\documentclass[
    superscriptaddress,
    prb,
    preprintnumbers,
    twocolumn,	%
    final,		%
    nobalancelastpage,
    nonbibnotes,
    showpacs,
    fleqn,
    a4paper,
    showkeys
]{revtex4-1}

\newif\iftwocolumn
\twocolumntrue

\def\figAsize{0.45\textwidth}
\def\figBsize{60mm}
\def\figCsize{0.45\textwidth}

\def\figEsize{70mm}
\usepackage[cp1250]{inputenc}
\usepackage{graphicx}
\usepackage{epstopdf}
\usepackage{subfigure}
\usepackage{amsmath}
\usepackage{mathtools}
\usepackage{dcolumn}
\usepackage{bm}
\usepackage{soul}

\DeclareMathOperator\FFT{FFT}
\DeclareMathOperator\myRe{Re}
\DeclareMathOperator\myIm{Im}

\begin{document}

\title{Ferroelectric domain patterns in barium titanate single crystals studied by means of digital holographic microscopy}

\author{Pavel Mokr\'{y}}
\email{mokry@ipp.cas.cz}
\affiliation{Regional Centre for Special Optics and Optoelectronic Systems (TOPTEC), Institute of Plasma Physics, Academy of Sciences of the Czech Republic, Za Slovankou 1782/3, 182 00 Prague 8, Czech Republic}

\author{Pavel Psota}
\affiliation{Regional Centre for Special Optics and Optoelectronic Systems (TOPTEC), Institute of Plasma Physics, Academy of Sciences of the Czech Republic, Za Slovankou 1782/3, 182 00 Prague 8, Czech Republic}

\author{Kate\v{r}ina Steiger}
\affiliation{Regional Centre for Special Optics and Optoelectronic Systems (TOPTEC), Institute of Plasma Physics, Academy of Sciences of the Czech Republic, Za Slovankou 1782/3, 182 00 Prague 8, Czech Republic}

\author{Jan V\'{a}clav\'{\i}k}
\affiliation{Regional Centre for Special Optics and Optoelectronic Systems (TOPTEC), Institute of Plasma Physics, Academy of Sciences of the Czech Republic, Za Slovankou 1782/3, 182 00 Prague 8, Czech Republic}

\author{Roman Dole\v{c}ek}
\affiliation{Regional Centre for Special Optics and Optoelectronic Systems (TOPTEC), Institute of Plasma Physics, Academy of Sciences of the Czech Republic, Za Slovankou 1782/3, 182 00 Prague 8, Czech Republic}

\author{David V\'{a}penka}
\affiliation{Regional Centre for Special Optics and Optoelectronic Systems (TOPTEC), Institute of Plasma Physics, Academy of Sciences of the Czech Republic, Za Slovankou 1782/3, 182 00 Prague 8, Czech Republic}

\author{V\'{\i}t L\'{e}dl}
\affiliation{Regional Centre for Special Optics and Optoelectronic Systems (TOPTEC), Institute of Plasma Physics, Academy of Sciences of the Czech Republic, Za Slovankou 1782/3, 182 00 Prague 8, Czech Republic}

\date{\today}

\begin{abstract}
In this Article, we report on the observation of ferroelectric domain pattern in the whole volume of the ferroelectric barium titanate single crystal by means of the digital holographic microscopy (DHM).
Our particular implementation of DHM is based on the Mach-Zehnder interferometer and the numerical processing of data employs the angular spectrum method.
A modification of the DHM technique, which allows a fast and accurate  determination of the domain walls, i.e. narrow regions separating the antiparallel domains, is presented. 
Accuracy and sensitivity of the method are discussed.
Using this approach, the determination of important geometric parameters of the ferroelectric domain patterns (such as domain spacing or the volume fraction of the anti-parallel domains) is possible.
In addition to the earlier DHM studies of domain patterns in lithium niobate and lithium tantalate, our results indicate that the DHM is a convenient method to study a dynamic evolution of ferroelectric domain patterns in all perovskite single crystals. 

\end{abstract}

\keywords{
ferroelectric domain patterns;
electro-optical materials;
digital holographic microscopy
}

\maketitle

\section{Introduction}
\label{sec:intro}

It is known that macroscopic properties such as dielectric constant, piezoelectric coefficients, and elastic compliance of ferroelectric samples are tremendously enhanced by the existence of domain walls in the ferroelectric sample~\cite{fousek_open_2003}.
In the research field called domain engineering~\cite{fousek_domain_2001} this effect is employed in a controlled manner and several methods for the preparation of domain patterns to intentionally enhance particular physical properties of ferroelectric samples has been demonstrated so far~\cite{kalinin_ferroelectric_2004, li_controlled_2008, hou_laser-induced_2010}. 
An example of enhancement of piezoelectric properties in ferroelectrics using the domain engineering was presented in \cite{wada_enhanced_1999, wada_enhanced_2004-1, wada_enhanced_2005, wada_domain_2006, yako_enhanced_2006}.
Understanding the details of physical processes standing behind the aforementioned enhancements requires non-destructive methods for the observation of ferroelectric domain patterns, which allow a fast and accurate determination of their geometric properties.


Therefore, it is seen that the progress in domain engineering in ferroelectrics and recent improvements in dielectric characterization techniques  stimulate the development of methods that allow the three dimensional (3D) observation of ferroelectric domains and domain walls.
Since the ferroelectric domain walls and charged defects produce local variations of the electric and strain fields, they can be studied on the atomic scale by means of electron holography (EH) \cite{zhang_electron_1992,zhang_electron_1993}. 
Since the presence of the residual depolarizing field due to the uncompensated surface charges strongly affects the EH patterns, their interpretation requires a comprehensive and careful analysis including the precise knowledge of  electric boundary conditions of the observed sample \cite{cao_theory_1993}.
Recently, a Cherenkov second-harmonic generation (CSHG) has appeared as a powerful tool for the visualization of domain walls in ferroelectric bulk crystals with superb optical resolution~\cite{kampfe_optical_2014}.
Unfortunately, CSHG method allows the visualization of charged head-to-head domain walls only. 


A very efficient tool for a 3D observation of ferroelectric domains in lithium niobate (LN) using the Digital Holographic Interferometry (DHI) was demonstrated by Grilli et al.~\cite{grilli_real-time_2004}.
So far, several arrangements of the optical interferometer and several numerical methods for the processing of digital holograms have been used. 
The reflective grating interferometer to capture the digital hologram has been used by Grilli et al.~\cite{grilli_real-time_2004} and Paturso et al.~\cite{paturzo_investigation_2004}.
Several works~\cite{zhi_ultraviolet-infrared_2008, zhi_phase_2009, qu_quantitative_2006, qu_digital_2007, hou_laser-induced_2010, hou_-situ_2012} have demonstrated domain pattern visualization using the Mach-Zehnder interferometer.
The captured digital holograms have been processed using Fresnel diffraction formula approach~\cite{grilli_real-time_2004, paturzo_investigation_2004, paturzo_origin_2005, paturzo_2d_2006} or angular spectrum method~\cite{qu_quantitative_2006, qu_digital_2007, zhi_ultraviolet-infrared_2008}.
Unfortunately, most of the ferroelectric domain patters observations have been performed only on LN and lithium tantalate single crystal samples, which are characterized by large values of coercive fields, stable domain patterns, and small values of piezoelectric coefficients. 


In the work presented below, we demonstrate the applicability of the DHI for the observation of ferroelectric domains in the barium titanate (BT) single crystals.
An important task resolved in this work is the limitation on the maximum electric field applied to the BT single crystal samples due to much smaller values of the coercive field. 
The small values of the coercive field put requirements on the sensitivity of the implemented digital holographic microscope (DHM).
The main objective of the work is to observe the periodic 180${}^\circ$ domain pattern in the BT single crystal sample and to develop a simple and fast method for the accurate identification of the fundamental geometric parameters of the observed domain pattern, namely the domain spacing and the volume fractions of the anti-parallel ferroelectric domains.
In order to achieve the objective, a DHM based on the Mach-Zehnder interferometer has been constructed and the angular spectrum method is used for the numerical processing of the interferometric measurements.

\section{Methods}
\label{sec:Methods}

\begin{figure}[t]
\begin{center}
	\includegraphics[width=\figAsize]{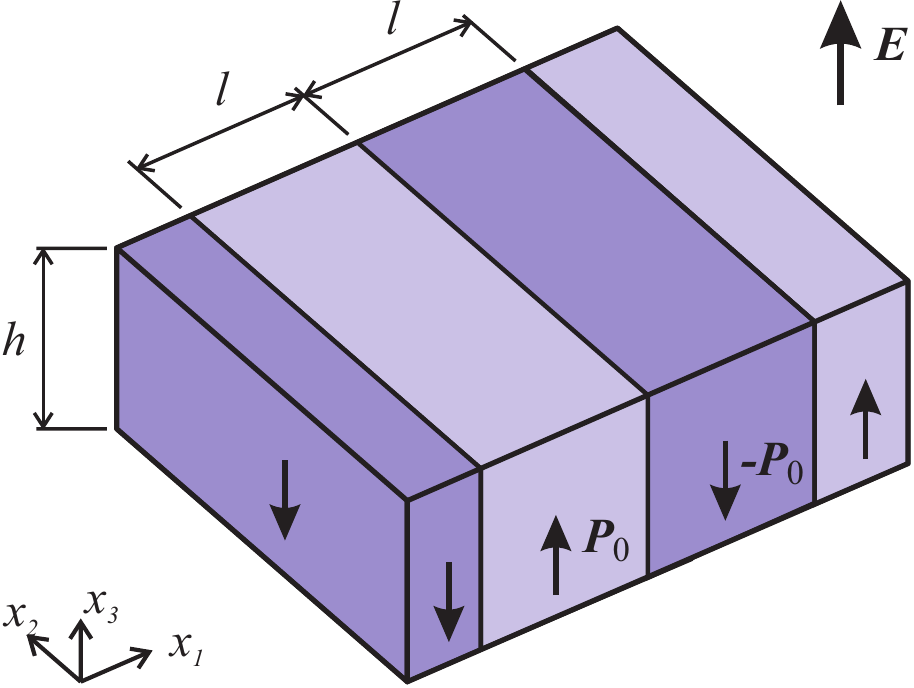}
\end{center}
\caption{
Geometry of the considered [001]-oriented ferroelectric single crystal of barium titanate. Vectors of spontaneous polarization in the adjacent ferroelectric domains are oriented in the direction of $x_3$ axis. Thickness of the ferroelectric sample is denoted by the symbol $h$. The periodic domain pattern with the half-period denoted by the symbol $l$ is considered.
}
\label{fig:geometry}
\end{figure}
%
In order to implement the methods for the DHI observation of ferroelectric domain patterns, it is necessary to specify basic requirements for the examined samples.
We consider a [001]-oriented ferroelectric single crystal of barium titanate, which is shown in Fig.~\ref{fig:geometry}.
Vectors of spontaneous polarization $P_0$ in the adjacent ferroelectric domains are oriented in the direction of $x_3$ axis.
It is considered that the periodic domain pattern with the half-period denoted by the symbol $l$ is formed in the ferroelectric sample and that all domain walls crosses the whole thickness $h$ of the sample.


The observation of ferroelectric domains using the optical interferometry is based on the employment of nonlinear polarization-optic effect, i.e. the relation between the refractive index tensor and the polarization of the crystal lattice.
This physical phenomenon can be expressed using the Taylor expansion of the optical indicatrix tensor in terms of the polarization vector components:
\begin{equation}
	\label{eq:indicatrixP}
	\left(\frac 1{n^2}\right)_{ij}(P)= \frac 1{n_0^2}\,\delta_{ij} + g_{ijkl}^T\,P_kP_l,
\end{equation}
where $n_0$ is the virtually temperature independent isotropic refractive index in the paraelectric phase of perovskites and $g_{ijkl}^T$ are the unclamped polarization-optic coefficients, and $P_i$ is the polarization vector. 
Let us consider the hard-ferroelectric approximation where the polarization response of the ferroelectric single crystal is given by the sum of the constant spontaneous polarization $P_{0,i}$ (whose orientation alters from domain to domain) and the linear dielectric response of the crystal lattice to the external electric field $E_i$:
\begin{equation}
    \label{eq:HardFerroelApprox}
	P_i = \pm P_{0,i} + \chi_{ij} E_j,
\end{equation}
where $\chi_{ij}$ is the dielectric susceptibility tensor of the crystal lattice.
When Eq.~(\ref{eq:HardFerroelApprox}) is substituted into Eq.~(\ref{eq:indicatrixP}), the optical indicatrix can be expressed as the function of the applied electric field $E_i$ and the spontaneous polarization $P_{0,i}$ within each domain:
\begin{equation}
	\label{eq:indicatrixE}
	\left(\frac 1{n^2}\right)_{ij}= \frac 1{n_0^2}\,\delta_{ij} + g_{ijkl}^T\,P_{0,k}P_{0,l} \pm 
    2\, g_{ijkl}^T\,P_{0,k}\chi_{lm}\, E_m.
\end{equation}


The last term on the right-hand side of Eq.~(\ref{eq:indicatrixE}) expresses the linear electro-optic effect, which is the source of the optical contrast, i.e. the difference in the refractive index $\Delta n_{ij}$, between the adjacent anti-parallel ferroelectric domains at a given constant applied electric field $E_k$:
\begin{equation}
	\label{eq:Deltanij}
	\Delta n_{ij}(E_k) = n_c^3 r_{ijk}^T E_k,
\end{equation}
where 
\begin{equation}
	\label{eq:rijk}
    r_{ijk}^T=g_{ijnm}^T\chi_{c,nk}P_{0,m}	
\end{equation}
is the unclamped linear electro-optic coefficient tensor and $n_c=n_0-(1/2)\, g_{11}^T\, n_0^3\, P_{0,3}^2$ stands for the value of the refractive index along the polar axis in barium titanate. 
The essential point, which is expressed by Eq.~(\ref{eq:rijk}), is the fact that the value of the linear electro-optic coefficient is proportional to the spontaneous polarization $P_0$.


In order to measure the spatial distribution of the refractive index in the ferroelectric single crystal, the ferroelectric sample is illuminated by a light beam along the direction of the polar axis $x_3$.
In the absence of the electric field in the ferroelectric sample, the wavefront of the optical wave passes through the ferroelectric sample without any distortion. 
In the presence of the electric field $E_3$, the wavefront becomes distorted with a phase shift $\Delta\phi$, which is spatially modulated as the optical waves passes through the crystal with anti-parallel domains. 
The phase shift consists of the sum of two contributions: due to the electro-optic effect  and due to the crystal elongation produced by the piezoelectric effect~\cite{zgonik_dielectric_1994}: 
\begin{equation}
	\label{eq:DeltaPhi}
	\Delta\phi = -\frac{\pi\,n_c^3}{\lambda_0}\,
    	\left[
        	r_{333}^T - \frac{2\,\left(n_c-n_g\right)}{n_c^3}\, d_{333}
        \right]\, h E_3.
\end{equation}
The symbol $\lambda_0$ stands for the wavelength of light in a vacuum, $n_c$ is the refractive index along the polar axis, $n_g$ is the refractive index of the surrounding medium, $d_{333}$ is the piezoelectric coefficient.


The spatial dependence of the wavefront distortion, i.e. the spatial dependence of the phase shift $\Delta\phi$, which corresponds to the spatial distribution of the spontaneous polarization within the ferroelectric domain pattern, can be measured using a digital holographic microscopy (DHM).
The contemporary conventional Charge-coupled device (CCD) or Complementary metal-oxide-semiconductor (CMOS) digital cameras allow to measure only the intensity of the incident optical wave, which corresponds to the amplitude of the wave.
Unfortunately, the phase component of the wave is lost during the measurement process.


In order to deal with this drawback of digital measurements and to obtain information about the phase profile of the transmitted optical wave, the superposition with an additional reference wave is used.
This approach is called the Digital Holography Interferometry (DHI) and consists, in general, of two steps.


In the first step, a micro-interference pattern produced by the superposition of the reference wave $U_{\rm ref}$ and a the wave $U_o$ reflected from (or transmitted through) the measured object (it will be further referred as the object wave) is captured using the CCD or CMOS camera.
It should be noted here that the functions $U_{\rm ref}$ and $U_o$ are treated as complex numbers where their absolute values and arguments correspond to the  values of the amplitudes and phases of particular optical waves. 
The digital camera captures the intensity of the superposition of the reference and object waves and the captured digital object is called the digital hologram $H$:
\begin{equation} 
	\label{eq:Hdef}
	H = |U_{\rm ref}+U_o|^2 = 
    |U_o|^2 +|U_{\rm ref}|^2 + U_{\rm ref}\,U_o^\star+ U_{\rm ref}^\star\,U_o.
\end{equation}


In the second step, the phase parameter of the object wave is extracted numerically from the digital hologram $H$ using fundamental properties of the wave propagation, interference and diffraction. 
Usually, the second step in the DHI is called the digital hologram reconstruction.


In order to capture a hologram, which allows the numerical reconstruction of the parameters of the object wave, the both interfering waves are supposed to be coherent with the same polarization direction.
The digitally recorded hologram is transferred to a computer as a matrix of numbers, which are proportional to the intensity given by the function $H$ in Eq.~(\ref{eq:Hdef}).
Since the propagation of optical fields is completely described by the diffraction theory, the numerical reconstruction of the optical field can be processed numerically as an array of complex numbers representing the amplitude and phase of the interfering waves.
In a comparison with the classical ``analog'' holography, the reconstruction process is done completely numerically in the case of DHI.


Let us consider that the digital camera is located in a distance $d$ from the ferroelectric sample, where the phase distribution is of the interest. (It will be further referred as the object plane.)
In order to obtain the phase distribution of the object wave, the recorded digital hologram $H$ is multiplied with a numerical model of the conjugated reference wave $U_{\rm ref}^\star$.
It is convenient to consider the simplest case of the planar reference wave of the unite amplitude, so that we can put $U_{\rm ref}^\star=1$ at the hologram plane.
The numerically reconstructed complex field $U$ at the object plane, which contains the phase information of the object wave, is calculated by the Sommerfeld formula which describes the diffraction of a light wave by the hologram grating in the distance $d$ from the hologram:
\iftwocolumn
\begin{multline} 
	\label{eq:Udef}
	U(x_1,\,x_2) = 
    \frac{1}{j\lambda_0} 
	\iint H(\xi_1, \xi_2)\, \times \\
    U_{\rm ref}^\star(\xi_1, \xi_2)\, 
    	\frac{\exp(jk\delta)}{\delta}\, d\xi_1 d\xi_2,
\end{multline}
\else
\begin{equation} 
	\label{eq:Udef}
	U(x_1,\,x_2) = 
    \frac{1}{j\lambda_0} 
	\iint H(\xi_1, \xi_2)\,   
    U_{\rm ref}^\star(\xi_1, \xi_2)\, 
    	\frac{\exp(jk\delta)}{\delta}\, d\xi_1 d\xi_2,
\end{equation}
\fi
where
\begin{equation*}
	\delta=\sqrt{d^2+(\xi_1-x_1)^2+(\xi_2-x_2)^2}.
\end{equation*}
The symbols $x_1$ and $x_2$ stand for the coordinates at the object plane. 
The symbols $\xi_1$ and $\xi_2$ stand for the coordinates at the hologram plane. 


The Sommerfeld integral can be solved by several techniques~\cite{kreis_handbook_2005}.
For the purpose of DHM, we use the Angular Spectrum (AS) method, since it provides valid diffraction fields even for small distances $d$. 
In the AS method, the reconstruction formula Eq.~(\ref{eq:Udef}) can be interpreted as a superposition integral:
\iftwocolumn
\begin{multline} 
	\label{eq:Umod}
	U(x_1,\,x_2)=
    \iint H(\xi_1, \xi_2)\,
    \times \\
    U_{\rm ref}^\star(\xi_1, \xi_2)\, 
    g(x_1-\xi_1,\,x_2-\xi_2)\, d\xi_1 d\xi_2,
\end{multline}
\else
\begin{equation} 
	\label{eq:Umod}
	U(x_1,\,x_2)=
    \iint H(\xi_1, \xi_2)\,
    U_{\rm ref}^\star(\xi_1, \xi_2)\, 
    g(x_1-\xi_1,\,x_2-\xi_2)\, d\xi_1 d\xi_2,
\end{equation}
\fi
where $g$ is the impulse response. 


In practice, the integral in Eq.~(\ref{eq:Umod}) is computed using Fast Fourier Transform (FFT) as follows:
\iftwocolumn
\begin{multline} 
	\label{eq:UFFT}
	U(x_1,\,x_2) = 
    	\FFT_{\nu_1,\,\nu_2}^{-1} \left\{ 
        	\FFT_{\xi_1, \xi_2} \left[
            	H(\xi_1, \xi_2)\, 
\right. \right.\times\\ \left.\left.
                U_{\rm ref}^\star(\xi_1, \xi_2)
            \right]\left(\nu_1,\,\nu_2\right)\, G(\nu_1,\,\nu_2)
        \right\}(x_1,\,x_2).
\end{multline} 
\else
\begin{equation} 
	\label{eq:UFFT}
	U(x_1,\,x_2) = 
    	\FFT_{\nu_1,\,\nu_2}^{-1} \left\{ 
        	\FFT_{\xi_1, \xi_2} \left[
            	H(\xi_1, \xi_2)\, 
                U_{\rm ref}^\star(\xi_1, \xi_2)
            \right]\left(\nu_1,\,\nu_2\right)\, G(\nu_1,\,\nu_2)
        \right\}(x_1,\,x_2).
\end{equation} 
\fi
\iftwocolumn
	\def\myinsert{\hspace{-45mm}& \\}
\else
	\def\myinsert{}
\fi
where the symbol $G$ represents the transfer function with spatial frequencies $\nu_1$ and $\nu_2$:
\begin{equation} 
	\label{eq:Gdef}
	G(\nu_1,\,\nu_2) = 
    	\begin{dcases*}
		    \exp \left[
		    	-\frac{2\pi d}{\lambda_0}\,
		        \sqrt{1 - (\lambda_0\nu_1)^2-(\lambda_0\nu_2)^2}
			\right] 
            \myinsert
            & for $(\lambda_0 \nu_1)^2+(\lambda_0 \nu_2)^2 \leq 1$,\\
			0 & 
			otherwise.
		\end{dcases*}
\end{equation}


In order to measure the spatial distribution of the phase shift $\Delta\phi(x_1,\,x_2)$ given by Eq.~(\ref{eq:DeltaPhi}), two digital holograms are captured. 
The first hologram is captured in the initial (reference) state of the ferroelectric sample at the absence of the electric field in the ferroelectric sample.
The reconstructed object complex wave $U_1$ in the reference state has the form:
\begin{equation} 
	\label{eq:U1def}
	U_1(x_1,\,x_2)=\left|U_1(x_1,\,x_2)\right|\,
    	\exp\left[j\phi_1(x_1,\,x_2)\right].
\end{equation}
The second hologram is captured when a non-zero external electric field $E_3$ is applied to the ferroelectric sample. 
The reconstructed object complex wave $U_2$ in this situation has the form:
\begin{equation} 
	\label{eq:U2def}
	U_2(x_1,\,x_2)=\left|U_1(x_1,\,x_2)\right|\,
    	\exp\left[j\phi_2(x_1,\,x_2)\right].
\end{equation}
Finally, the spatial distribution of the phase shift $\Delta\phi(x_1,\,x_2)$ is computed from the fraction $U_2/U_1$ using the formula:
\begin{equation} 
	\label{eq:DeltaPhiMeas}
	\Delta\phi = \arctan \left\{
    	\frac{\myIm\left(U_2\,U_1^\star\right)}{\myRe\left(U_2\,U_1^\star\right)} 
    \right\}
    = \arctan \left\{
    	\frac{
        	U_1^\prime\, U_2^{\prime\prime} - U_2^\prime\, U_1^{\prime\prime}
        }{
        	U_1^{\prime}\, U_2^{\prime} + U_1^{\prime\prime}\, U_2^{\prime\prime}
        } 
    \right\}.
\end{equation}
Such an approach has several advantages, which can be beneficially used during the interpretation of numerical data.

\section{Results}
\label{sec:Results}

\begin{figure*}[t]
	\centering
	\subfigure[]{
		\includegraphics[height=\figBsize]{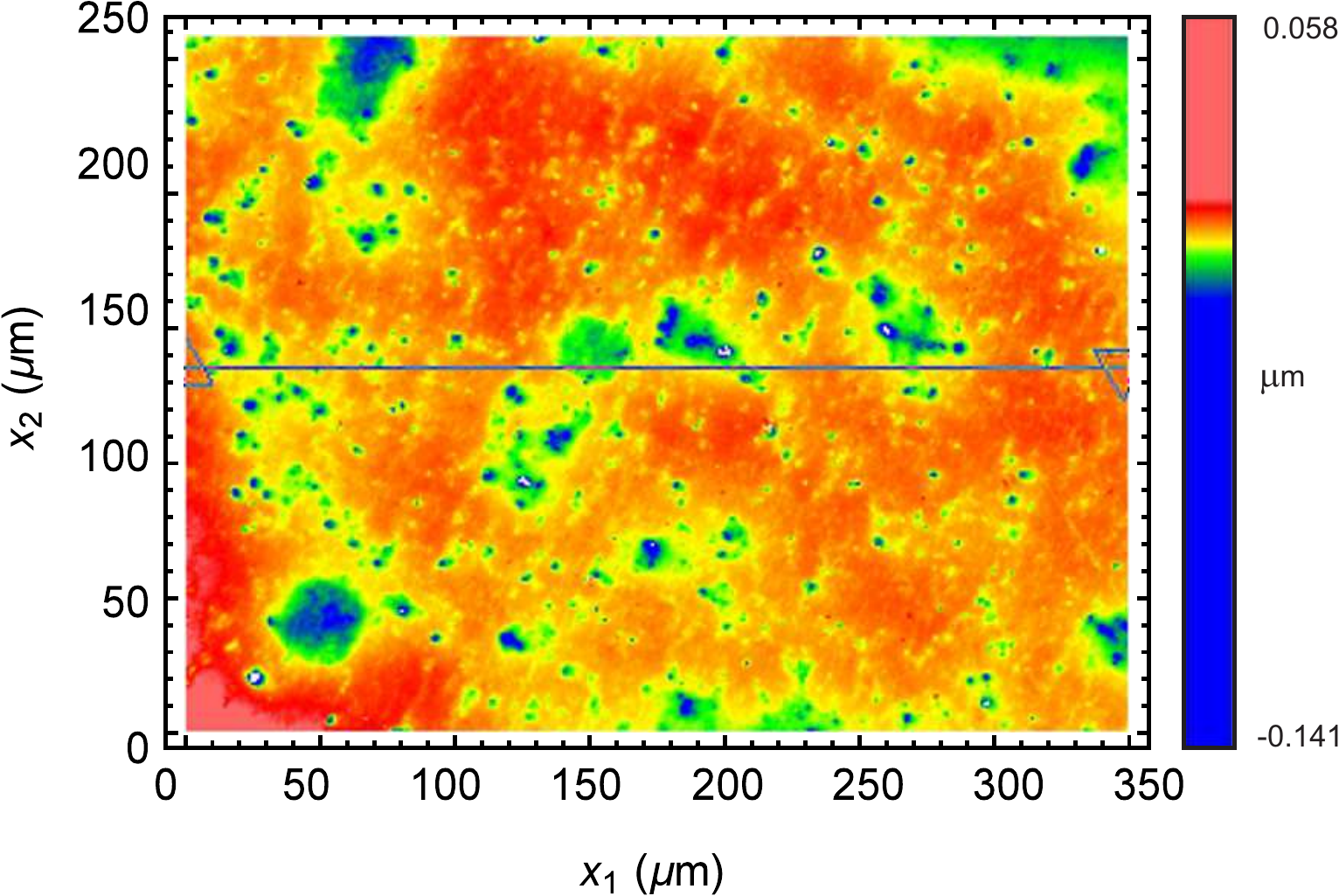}
		\label{fig:results:roughness:a}
	}
	\subfigure[]{
		\includegraphics[height=\figBsize]{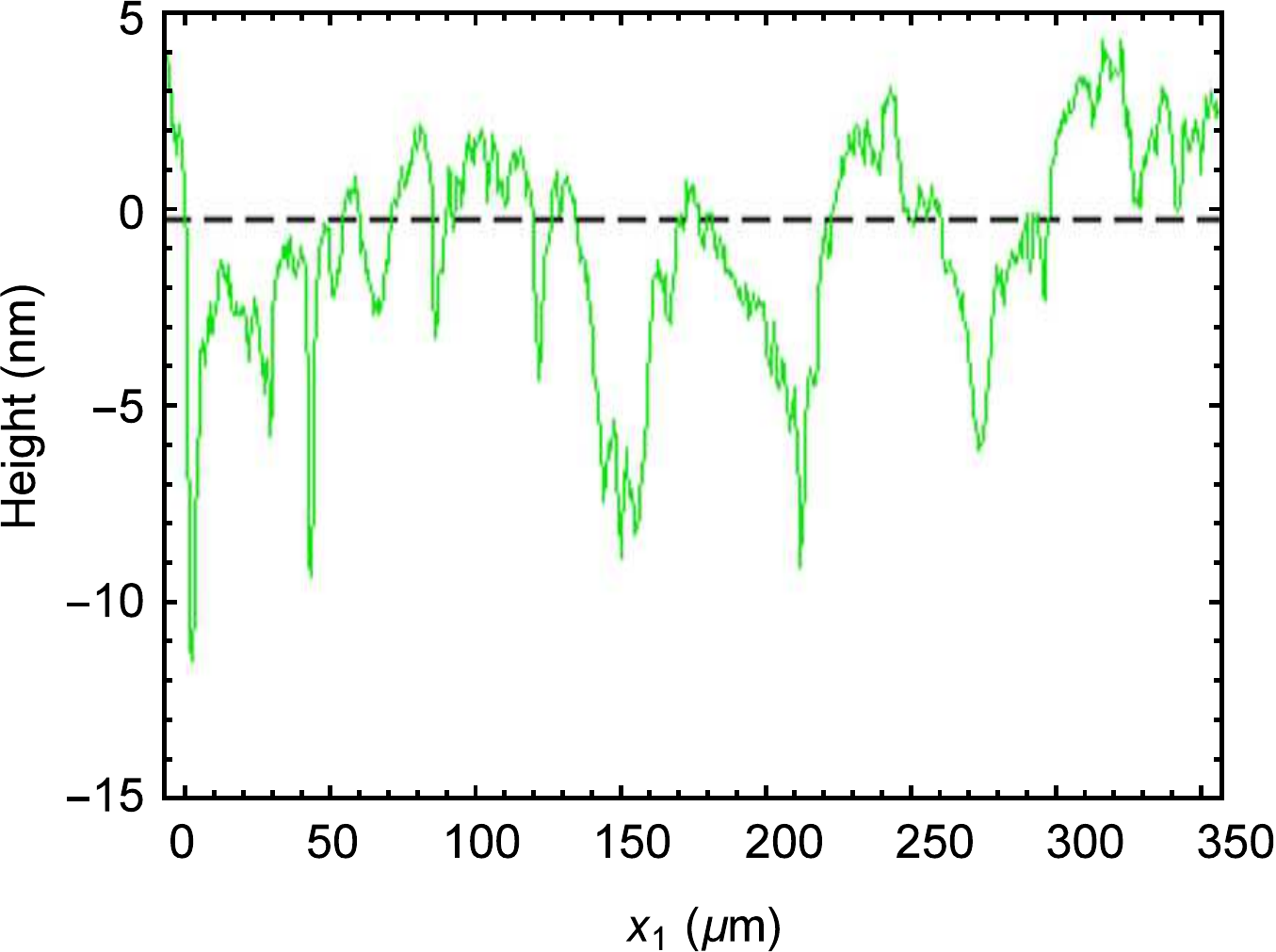}
		\label{fig:results:roughness:b}
	}
	\subfigure[]{
		\includegraphics[height=\figBsize]{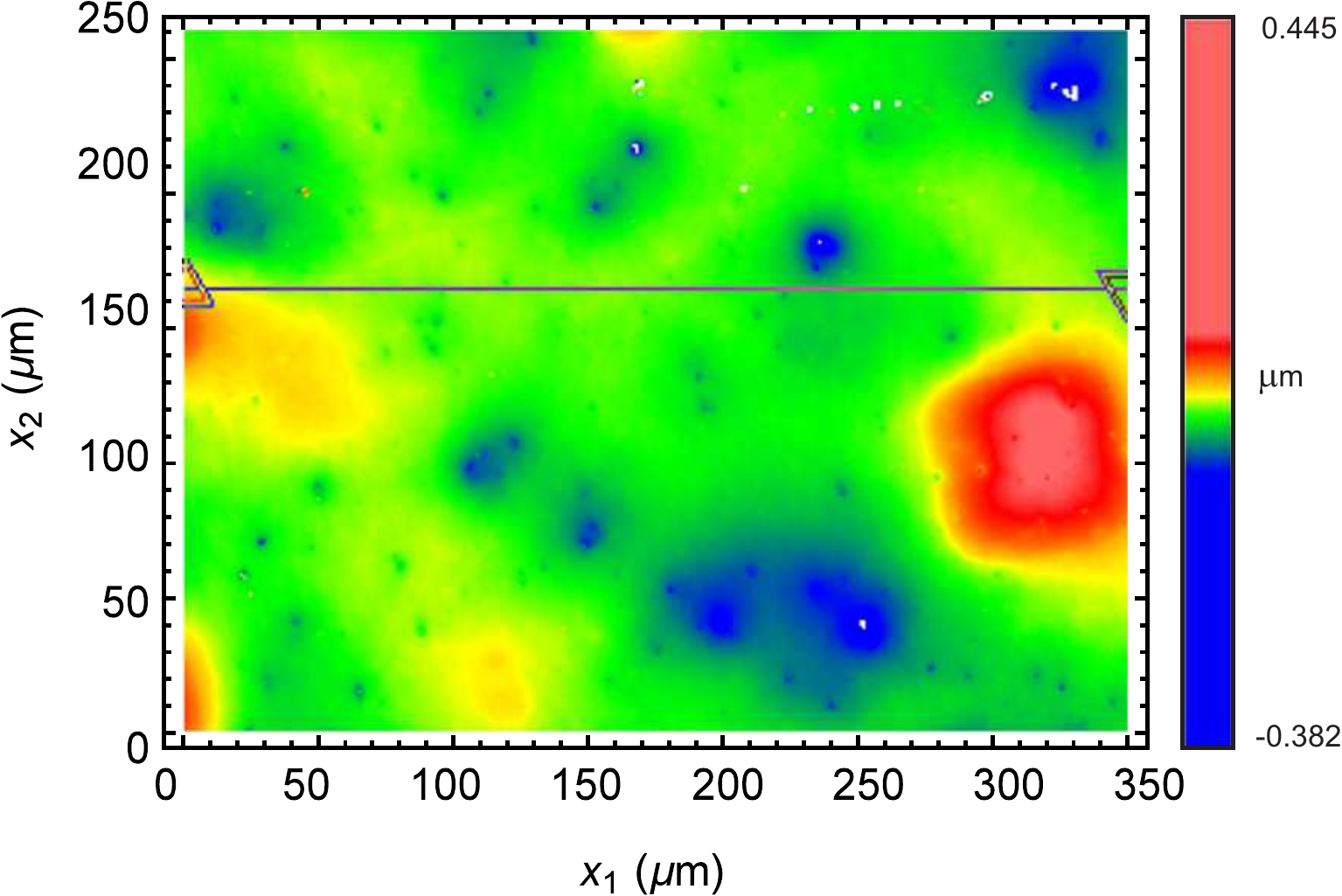}
		\label{fig:results:roughness:c}
	}
	\subfigure[]{
		\includegraphics[height=\figBsize]{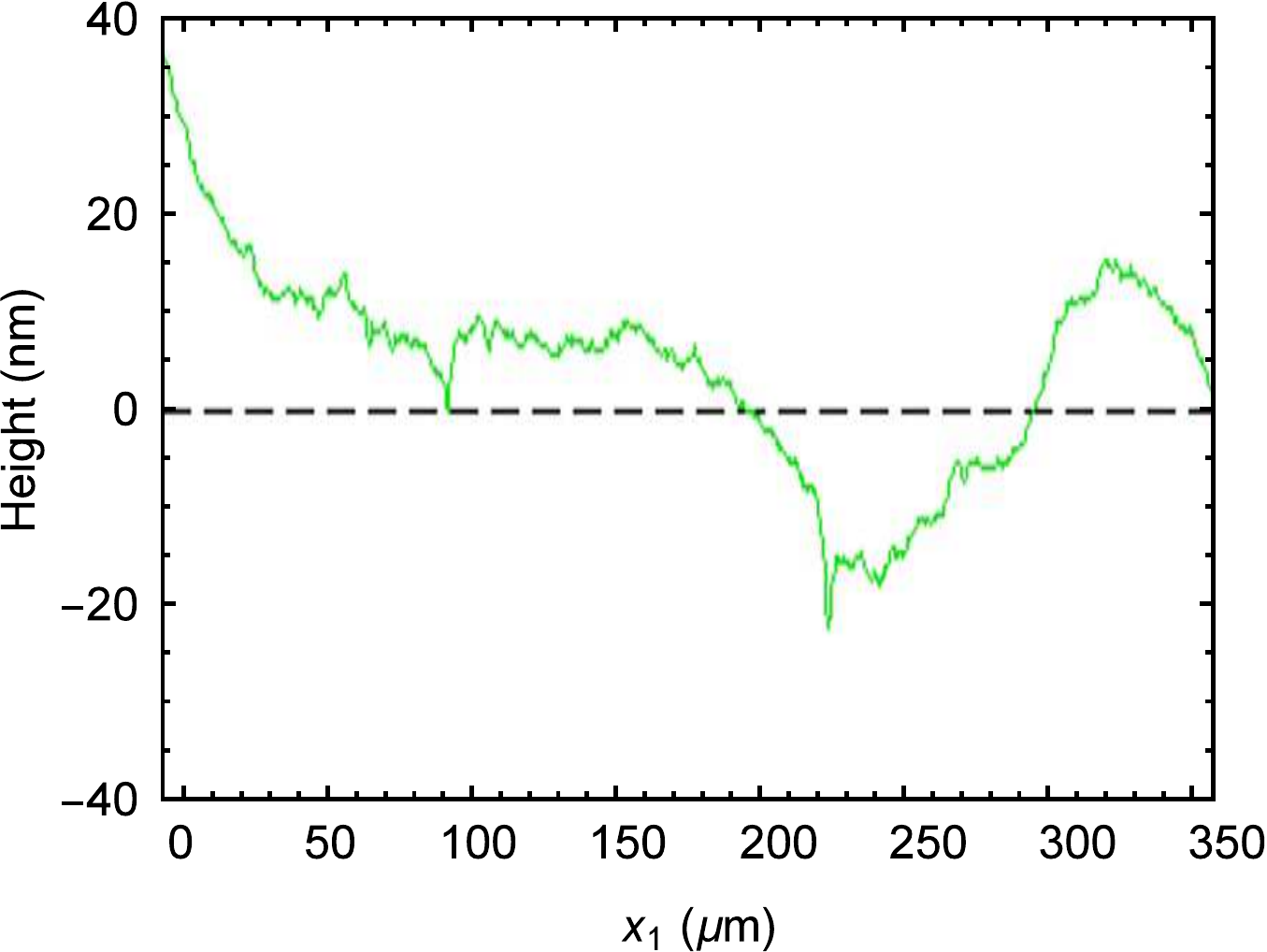}
		\label{fig:results:roughness:d}
	}
	\caption{
Roughness of the surface of the examined BT single crystal sample using white light interferometer (Zygo Co., USA). Figures \subref{fig:results:roughness:a} and \subref{fig:results:roughness:c} show the  roughness on the selected area of the opposite surfaces of the ferroelectric single crystal sample. Figures \subref{fig:results:roughness:b} and \subref{fig:results:roughness:d} show the roughness at the cross-sections indicated by lines in Figs.~\subref{fig:results:roughness:a} and \subref{fig:results:roughness:c}, respectively.
}
	\label{fig:results:roughness}
\end{figure*}
In our experiments, the [001]-oriented single crystals of pure barium titanate of the thickness $h=0.5$~mm (Forschunginstitut f\"ur mineralische und metallische Werkstoffe Edelsteine/Edelmetalle GmbH., Germany) have been examined.
In order to obtain a high quality optical surface, the BT samples have been polished in two steps. 
At first, the surface has been lapped with a free silicon carbide abrasive with a grain size 18~$\rm\mu$m and then polished with cerium oxide on the polyurethane substrate.
In order to provide the external voltage to the bulk of the ferroelectric single crystal, the optical cell with sapphire windows and liquid transparent electrodes (solution of barium chloride, BaCl$_2$, in water with the mass concentration of approx. 5~g/l) has been fabricated.
The surface roughness of the sample has been measured using the white light interferometer (NewView 7200, Zygo Co., USA equipped with 20$\times$ Mirau type lens with numerical aperture equal to 0.4) in the intermediate stage of the DHI experiments.
Figure~\ref{fig:results:roughness} shows the results of the surface roughness measurements. 
The root mean square values of surface roughness of 4.3 and 24.3~nm were measured.
These values are much smaller than the wavelength of the used laser source.
The crystallization of the BaCl$_2$ nuclei from its water solution is expected to be the main source of the measured surface roughness.


\begin{figure}
	\centering
	\includegraphics[width=\figCsize]{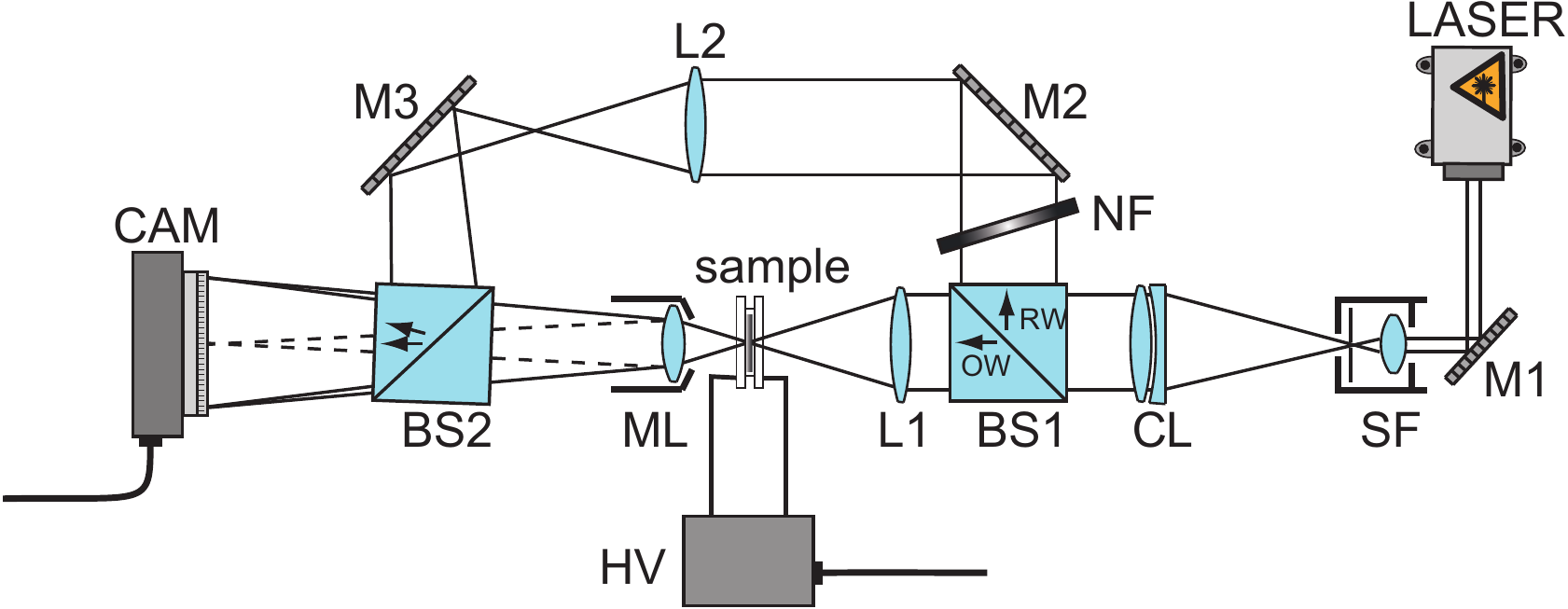}
\caption{Scheme of the constructed experimental setup for the three-dimensional mesoscopic observations of ferroelectric domain walls using the digital holographic microscope. The microscope consists of laser source (Laser), three mirrors (M1, M2, and M3), two beam splitters (BS1 and BS2), a spatial filter (SF), collimating lens (CL), two lens (L1 and L2), microscopic lens (ML), neutral density filter (NF), and digital camera (CAM). The microscope has the Mach-Zehnder arrangement. As the external voltage provided by the high-voltage source (HV) is applied to the ferroelectric single crystal liquid transparent electrodes, the phase of the  impinging beam (object wave, OW) is changed due to the linear electro-optic effect in the ferroelectric domains. The phase-modulated object wave then interferes with the reference wave (RW) and the resulting micro-interference pattern is captured by the digital camera as a digital hologram.}
\label{fig:DHMsetup}
\end{figure}
Fig.~\ref{fig:DHMsetup} shows the scheme of the constructed digital holographic microscope (DHM) in the transmission configuration. 
The arrangement is based on the Mach-Zehnder interferometer. 
The head of a Sapphire single frequency laser emits a beam having the wavelength of 488~nm and the power of 50~mW.
Behind the mechanical shutter, the beam is spatially filtered (SF), collimated (CO) and split by the polarizing beam splitter (BS1), equipped with half wavelength retardation plates in two beams.
Half wavelength retardation plates are used to set the intensity and polarization of each beam.
The first beam acts as a reference wave (RW) and could be further attenuated if necessary by a set of gray filters placed in filter wheels. 
The second beam called object wave (OW) is condensed by the lens (L1) and transmitted through the BT single crystal sample.


\begin{figure}
	\centering
	\includegraphics[width=\figCsize]{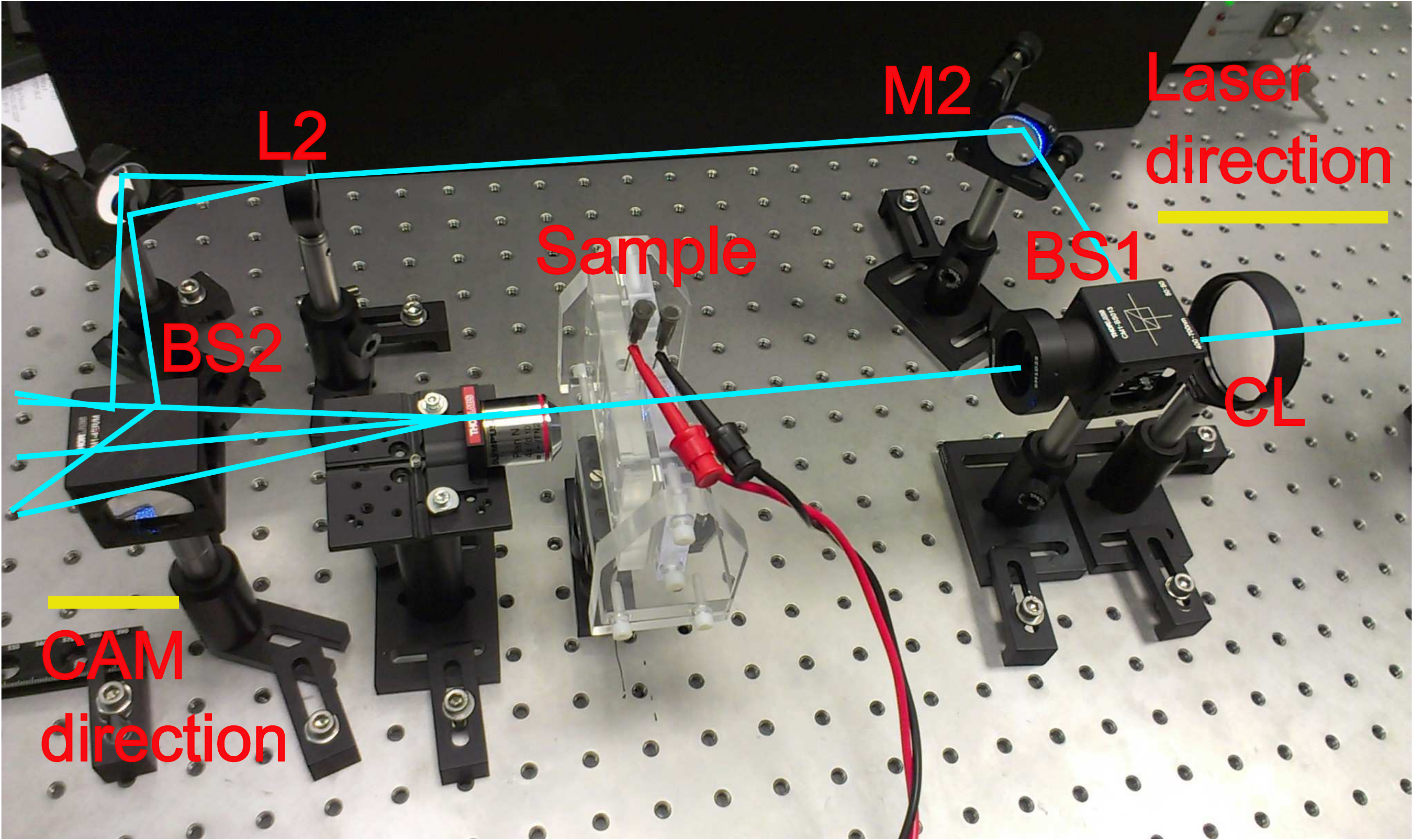}
\caption{Photograph of the constructed digital holographic microscope. Symbols standing by the optical elements are described in Fig.~\ref{fig:DHMsetup}.}
\label{fig:photo}
\end{figure}
The magnification of the image is achieved using the 4$\times$ Olympus plan achromat microscope lens (ML) with the numerical aperture of 0.10. 
The high numerical aperture of the ML reduces the spatial frequency of the sample in the image plane in order to fulfill the Nyquist criterion in the hologram plane and therefore the lateral resolution in the image plane can reach the diffraction limit. 
The L2 lens diverges the RW with the aim to match the curvature of the RW with the curvature of the OW in the hologram plane. 
The both waves are recombined using the beam splitter (BS2) and create an off-axis digital hologram that is captured by the IDS digital camera (CAM) with 2048 $\times$ 2048 pixels having the size 5.5~${\rm \mu}$m each. 
The position of the CAM (also called the hologram plane) is close to the position, where the image by the ML is formed. 
The photograph of the constructed device is shown in Fig.~\ref{fig:photo}.


\begin{figure*}[t]
	\centering
	\subfigure[$H_1$ at $E_3=0.0$~kV/mm]{
		\includegraphics[height=\figBsize]{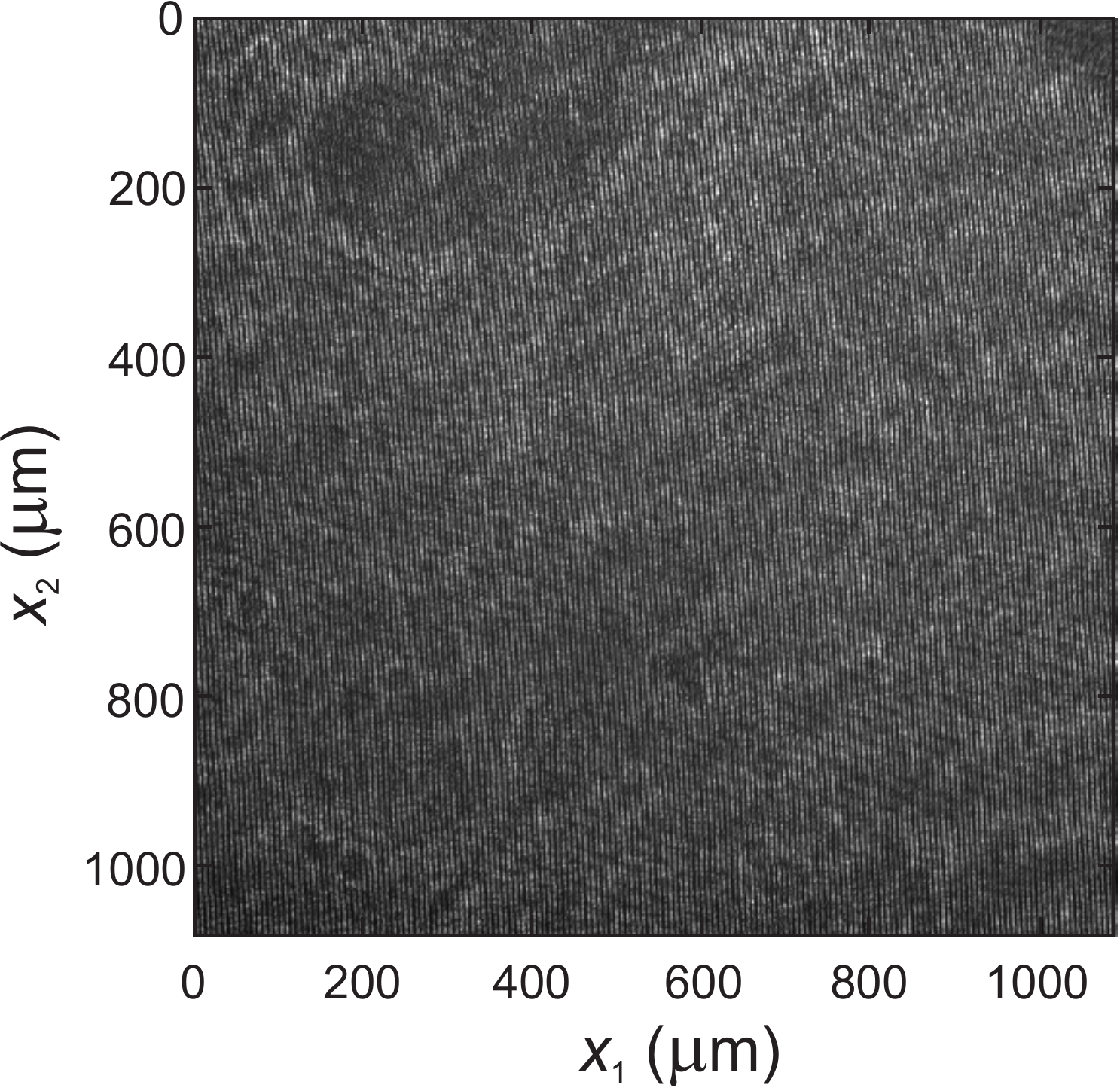}
		\label{fig:results:H:a}
	}
	\subfigure[$H_2$ at $E_3=0.8$~kV/mm]{
		\includegraphics[height=\figBsize]{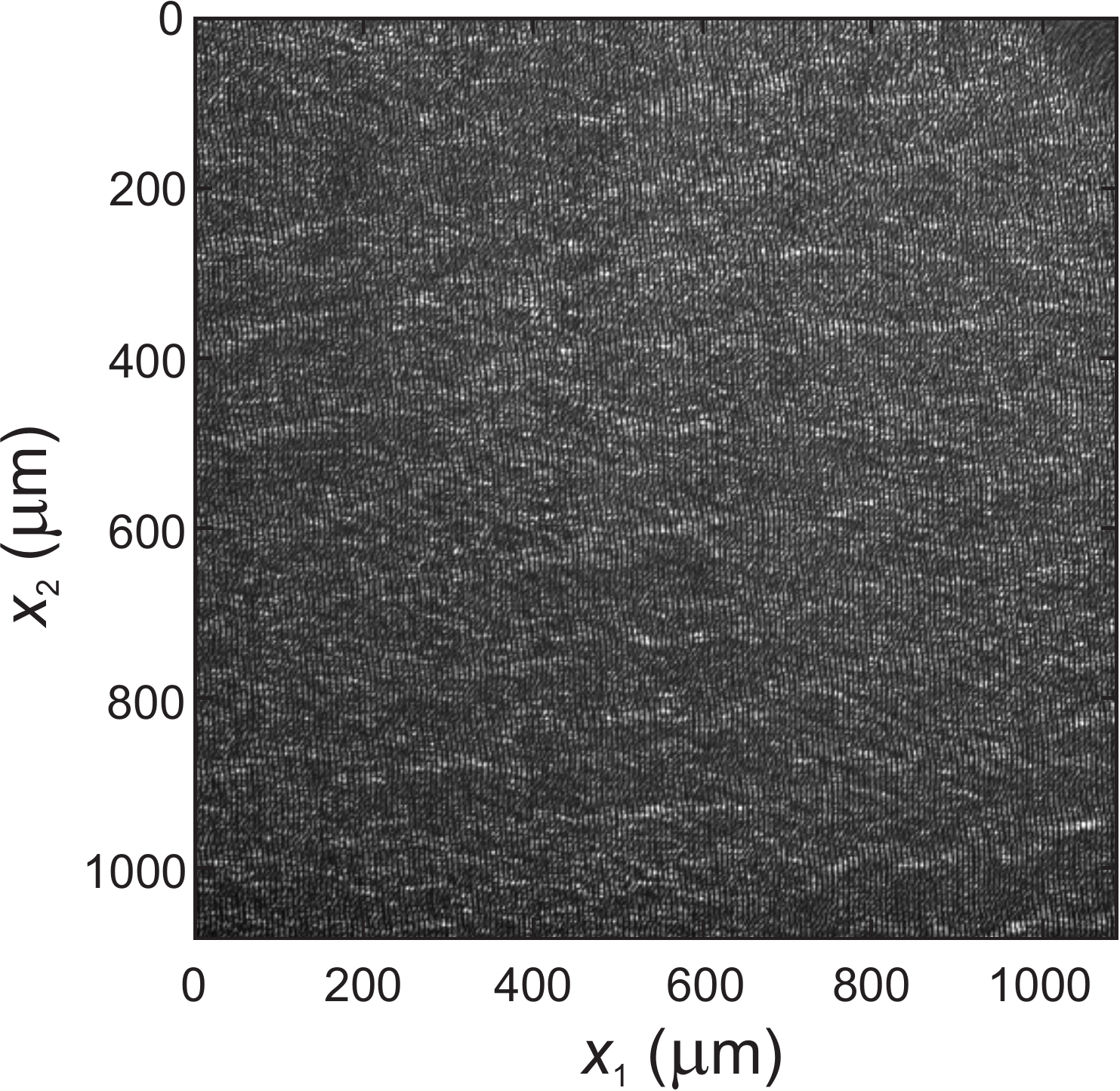}
		\label{fig:results:H:b}
	}
	\subfigure[$|U_1|$ at $E_3=0$~kV/mm]{
		\includegraphics[height=\figBsize]{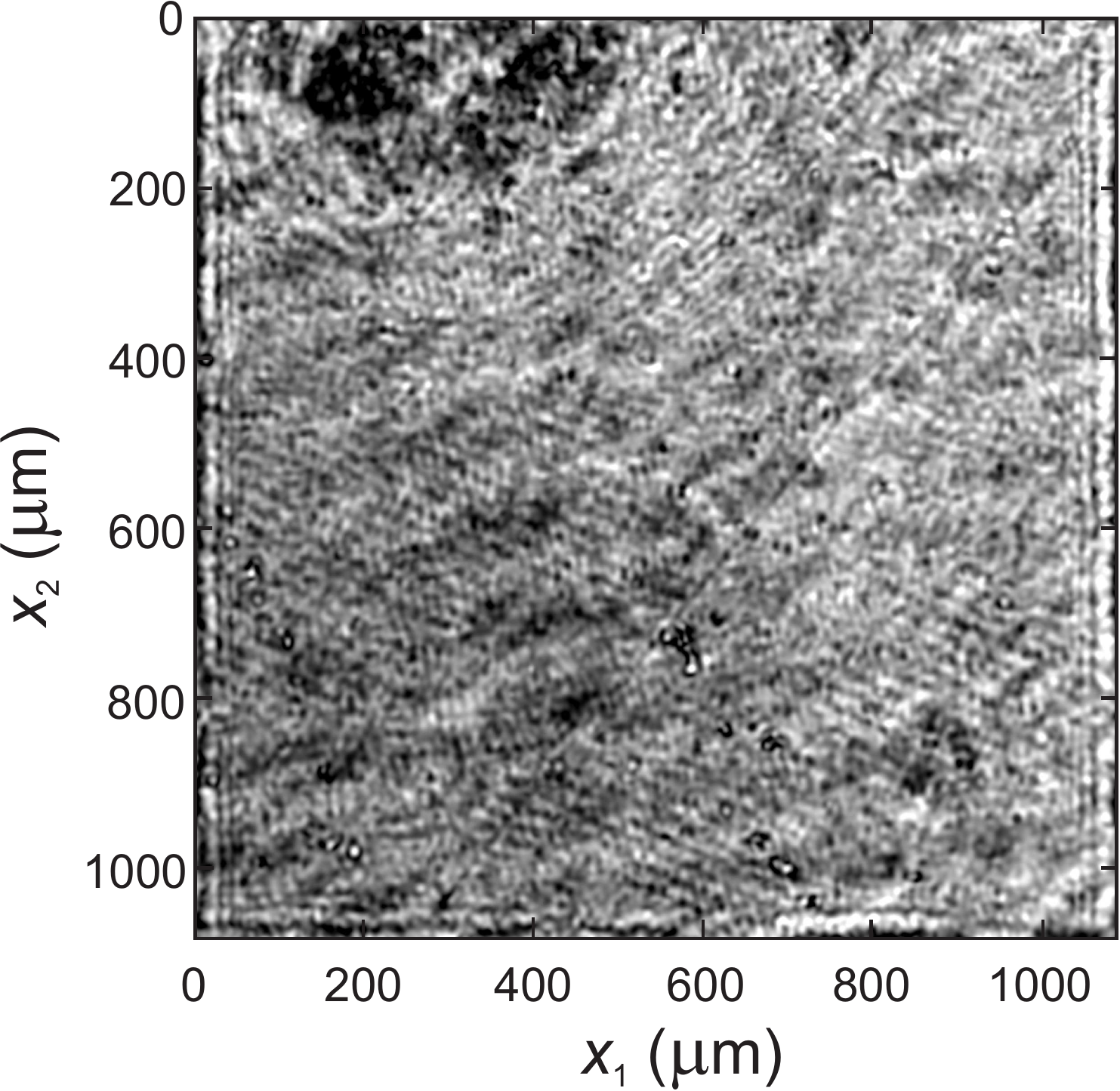}
		\label{fig:results:H:c}
	}
	\subfigure[$\arg U_1$ at $E_3=0$~kV/mm]{
		\includegraphics[height=\figBsize]{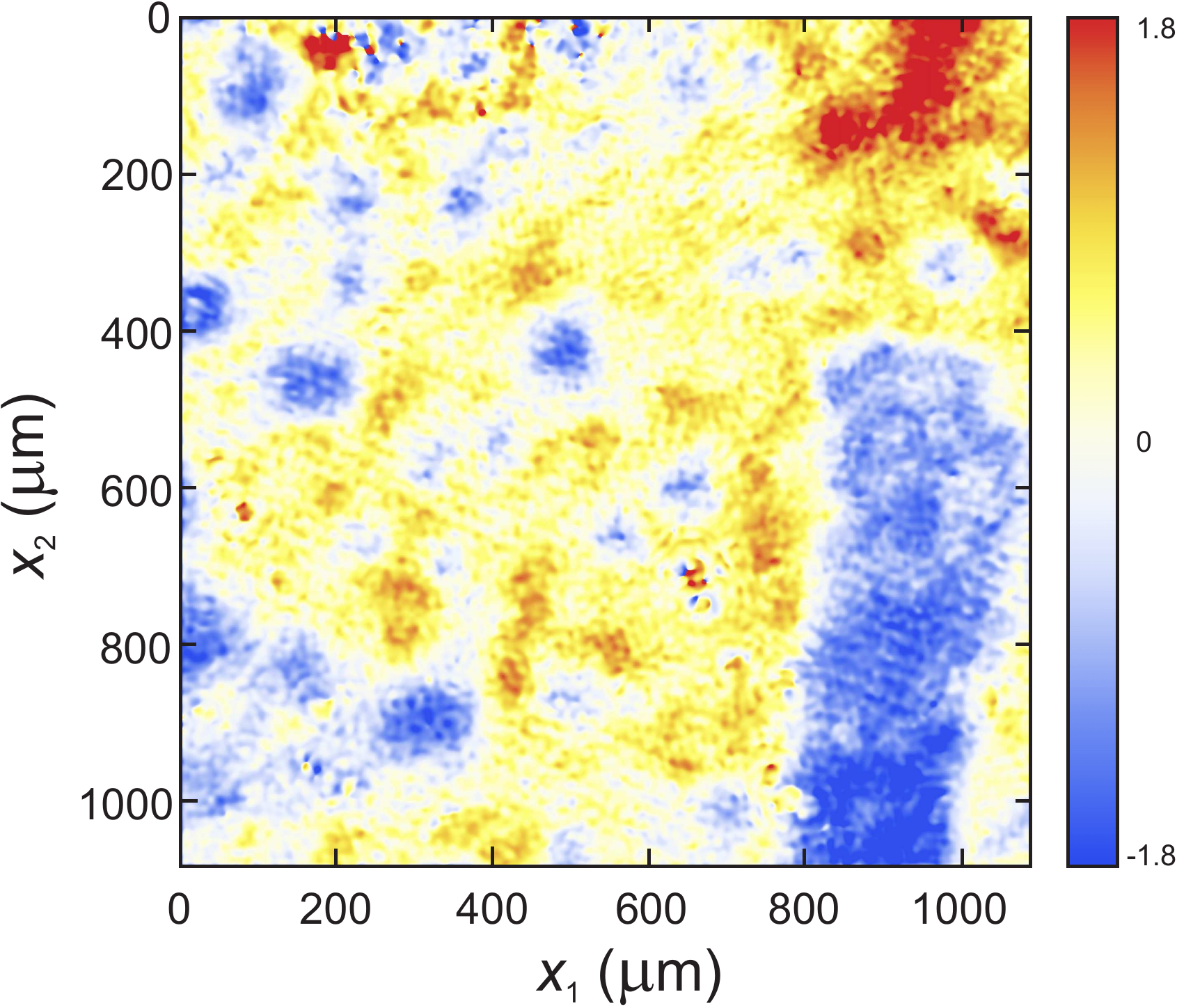}
		\label{fig:results:H:d}
	}
\caption{
\subref{fig:results:H:a} and \subref{fig:results:H:b} 
show micro-interference patterns called digital holograms $H_1$ and $H_2$ captured using digital camera during the observation of ferroelectric domain pattern in the BT single crystal at zero and $0.8$~kV/mm external electric field. The clear difference in the hologram patterns is produced due to the linear electro-optic effect. 
\subref{fig:results:H:c} shows the intensity of the optical wave transmitted through the observed volume of the BT single crystal.
\subref{fig:results:H:d} shows the computed wavefront deformation of the optical wave transmitted through the BT single crystal at zero applied field.
}
\label{fig:results:H}
\end{figure*}

\begin{figure}
	\centering
	\includegraphics[height=\figEsize]{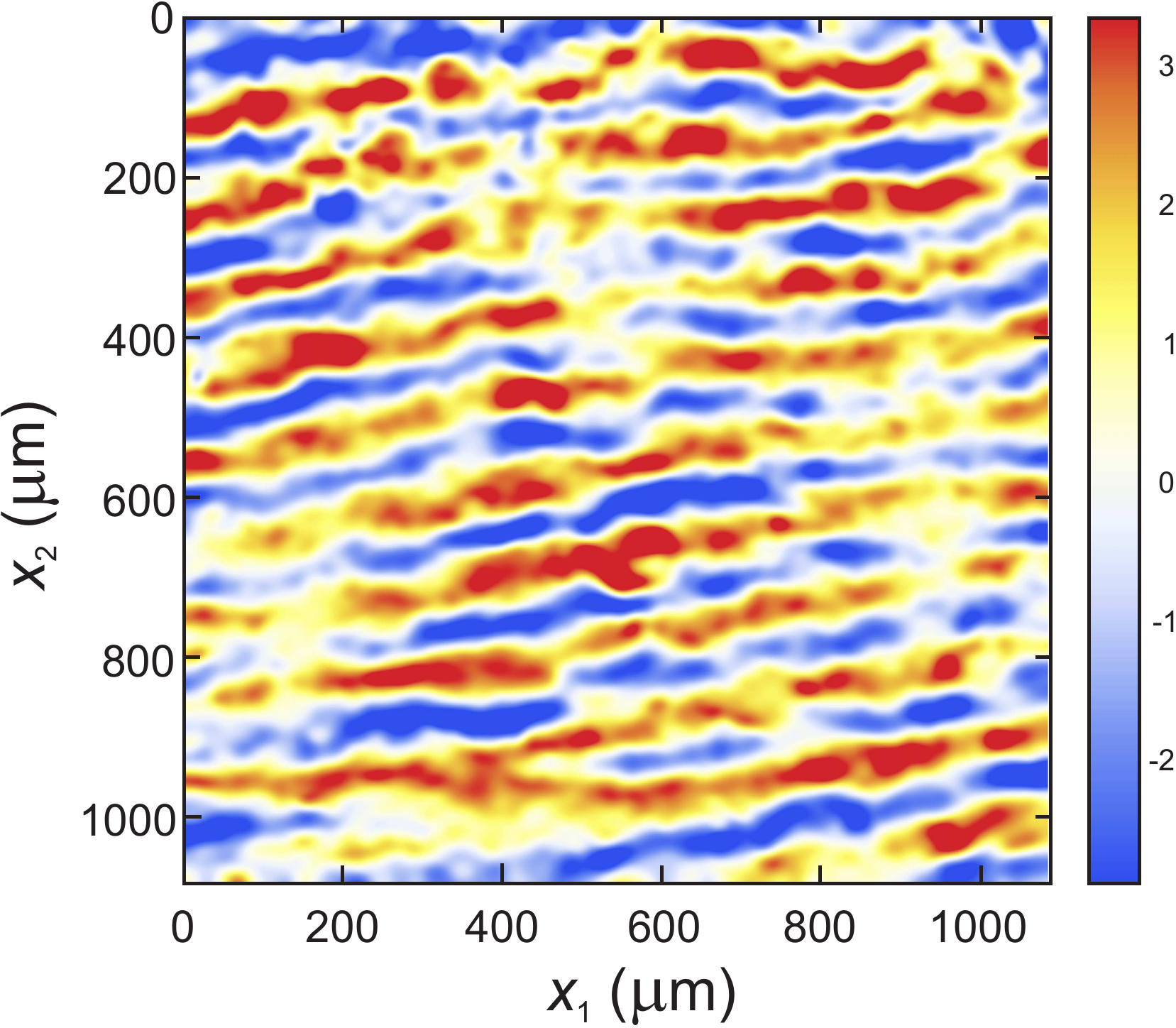}
\caption{
Observed distortion of the wavefront $\Delta\phi$ (see Eq.~\ref{eq:DeltaPhi}, in rad) of the light wave transmitted through the 180${}^\circ$ domain pattern in the [001]-oriented single crystal of barium titanate. The applied field of 0.8~kV/mm has been applied across the sample thickness.}
\label{fig:results:DeltaPhi}
\end{figure}
Figures~\ref{fig:results:H:a} and \ref{fig:results:H:b} show the captured digital holograms $H_1$ at zero electric field (Fig.~\ref{fig:results:H:a}) and $H_2$ at $E_3=0.8~$kV/mm (Fig.~\ref{fig:results:H:b}).
Each of the holograms $H_1$ and $H_2$ is used for the computation of the corresponding  complex wave $U_1$ and $U_2$, respectively, according to the procedure described by Eq.~(\ref{eq:UFFT}).
Figure~\ref{fig:results:H:c} shows the intensity of the optical wave $U_1$ transmitted through the observed volume of the crystal in the reference state. 
The intensity of the transmitted optical wave shows no evidence of the presence of 90${}^\circ$ domain walls, which would be otherwise clearly indicated by bright and dark stripes due to different orientations of the refractive index ellipsoid in the adjacent domains.
Figure~\ref{fig:results:H:c} shows the phase of the transmitted optical wave $U_1$, which is computed from the digital hologram $H_1$.
The phase profile of the complex field $U_1$ gives information on the spatial distribution of the refractive index in the observed volume of the BT single crystal at the zero applied field (with respect to the complex field in air). 
The difference between maximum and minimum values of the phase of $U_1$ is about 3.6~rad, which clearly indicates a great optical homogeneity of the observed part BT single crystal. 
The spatial variation of the refractive index is smaller than $10^{-5}$, which provides the clear evidence that the observed part of the BT single crystal is either the a-domain or the c-domain. 

The the distorted profile of the wavefront due to the anti-parallel domain pattern in the [001]-oriented BT single crystal is then computed from the known complex fields $U_1$ and $U_2$ using Eq.~(\ref{eq:DeltaPhiMeas}).
The resulting profile is shown in Fig.~\ref{fig:results:DeltaPhi}.

\section{Discussion} 
\label{sec:Discussion}

At the beginning of the discussion, it is necessary to determine what type of a domain pattern is observed using the DHI experiment.
Since the observed part of the BT single crystal is either the a-domain or the c-domain, the phase difference $\Delta\phi$, which is produced by the application of the external field due to the linear electro-optic effect, indicates the presence of 180${}^\circ$ domains.
%
%
%
%
Since the 180${}^\circ$ domains in the a-domain are undetectable by the linear electro-optic effect produced by the external electric field along the $x_3$-axis, the only possibility for the observed phase difference $\Delta\phi$ remains that the observed part of the crystal is in the c-domain, where the vector of spontaneous polarization is oriented along the $x_3$-axis.


In the next step, it is necessary to check whether the observed distortions of the transmitted wavefront are in an agreement with material parameters of BT available in literature. 
Figure~\ref{fig:results:DeltaPhi} shows that the application of the external electric field of 0.8~kV/mm has produced the phase shift by -2.94~rad in the areas on the sample surface, where the spontaneous polarization in the ferroelectric domain was oriented along the applied electric field (indicated by blue color in Fig.~\ref{fig:results:DeltaPhi}). 
In the areas corresponding to anti-parallel domains, the phase shift by 3.32~rad has been measured (indicated by red color in Fig.~\ref{fig:results:DeltaPhi}).
Let us consider the following numerical values, which are available in literature~\cite{zgonik_dielectric_1994} for BT single crystal: $n_c=2.36$, $d_{333}=90\times 10^{-12}$~C/N, and $r_{333}^T=105\times 10^{-12}$~m/V.
In addition, the following numerical values are considered in our experiment: $h=0.5$~mm, $\lambda_0=488$~nm, $n_g=1.33$, and $E_3=0.8$~kV/mm.
When the aforementioned values are substituted into Eq.~(\ref{eq:DeltaPhi}), the theoretical peak value of the phase shift is 3.08~rad. 
It means that the relative difference between measured and theoretical values are smaller that 5\%.
This indicates a reasonably good agreement between measured data and the available numerical values.


\begin{figure}
	\centering
	\includegraphics[height=\figEsize]{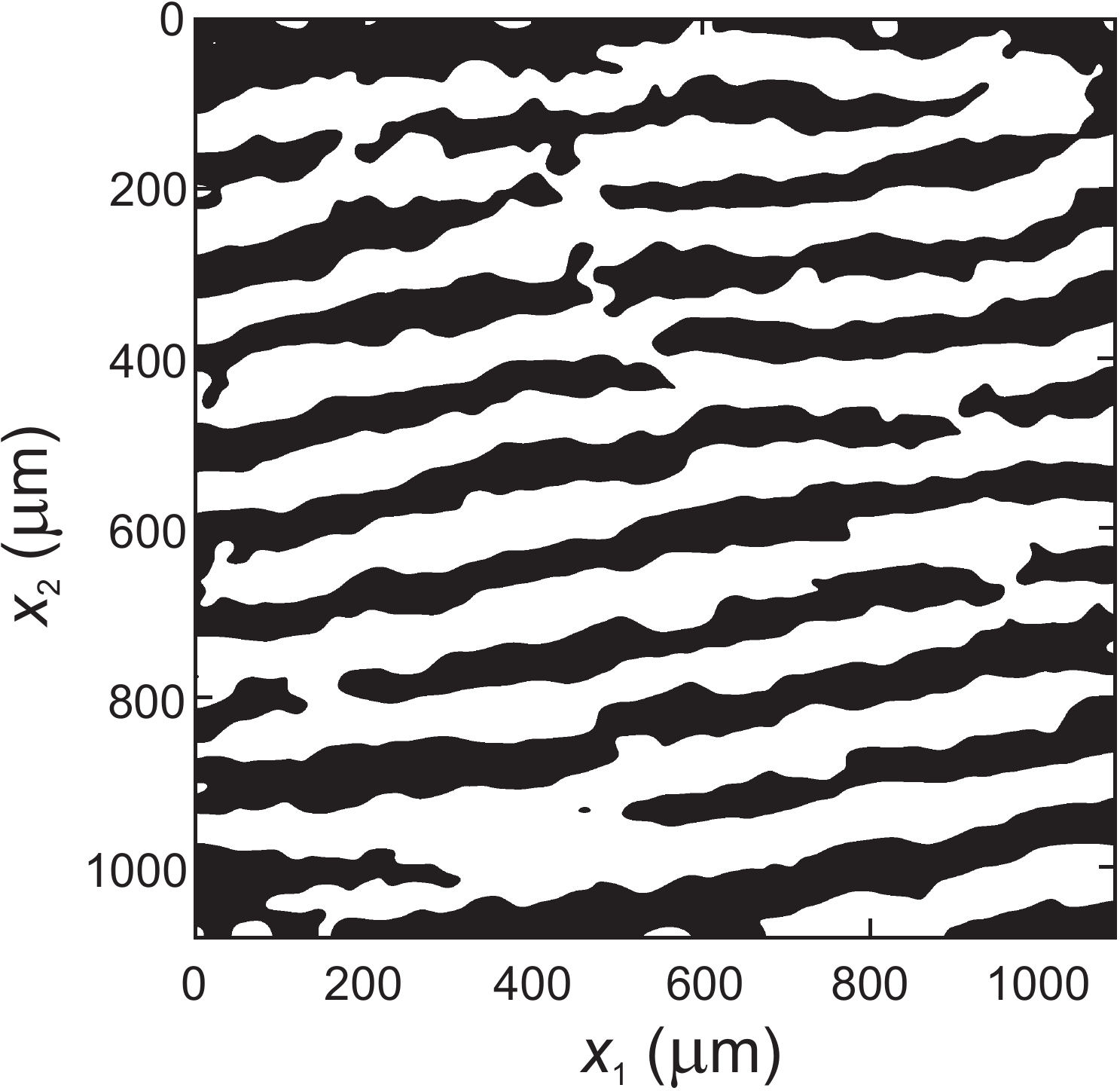}
\caption{Visualized 180${}^\circ$ domain pattern computed from the phase difference shown in Fig.~\ref{fig:results:DeltaPhi}. Black and white color indicates the domain with the vector of spontaneous polarization oriented along and against the applied electric field, respectively.}
\label{fig:results:Domains}
\end{figure}
The particular implementation of the DHM and the numerical processing of the captured digital holograms according to Eq.~(\ref{eq:DeltaPhi}) has a very convenient feature, which is the fact that the narrow regions, where approximately a zero-rad phase shift is computed, correspond to the cross-section of the domain walls and the sample surface.
This property of our approach makes it possible to easily locate the domain walls and distinguish the domains with the vector of spontaneous polarization oriented along (i.e. $P_0$) and against (i.e. $-P_0$) the applied electric field using a simple criterion of negative and positive value of the computed phase shift $\Delta\phi$, respectively.


Figure~\ref{fig:results:Domains} shows the visualization of the 180${}^\circ$ domain pattern computed from the phase difference shown in Fig.~\ref{fig:results:DeltaPhi} using the procedure described in the above paragraph. 
The results in Fig.~\ref{fig:results:Domains} indicate a clear plate-like domain pattern with the average value of the domain spacing identified as $l=47.7~{\rm \mu}$m.
The observed specks are produced by the surface contamination from the liquid electrodes.


\begin{figure}
	\centering
	\includegraphics[width=\figAsize]{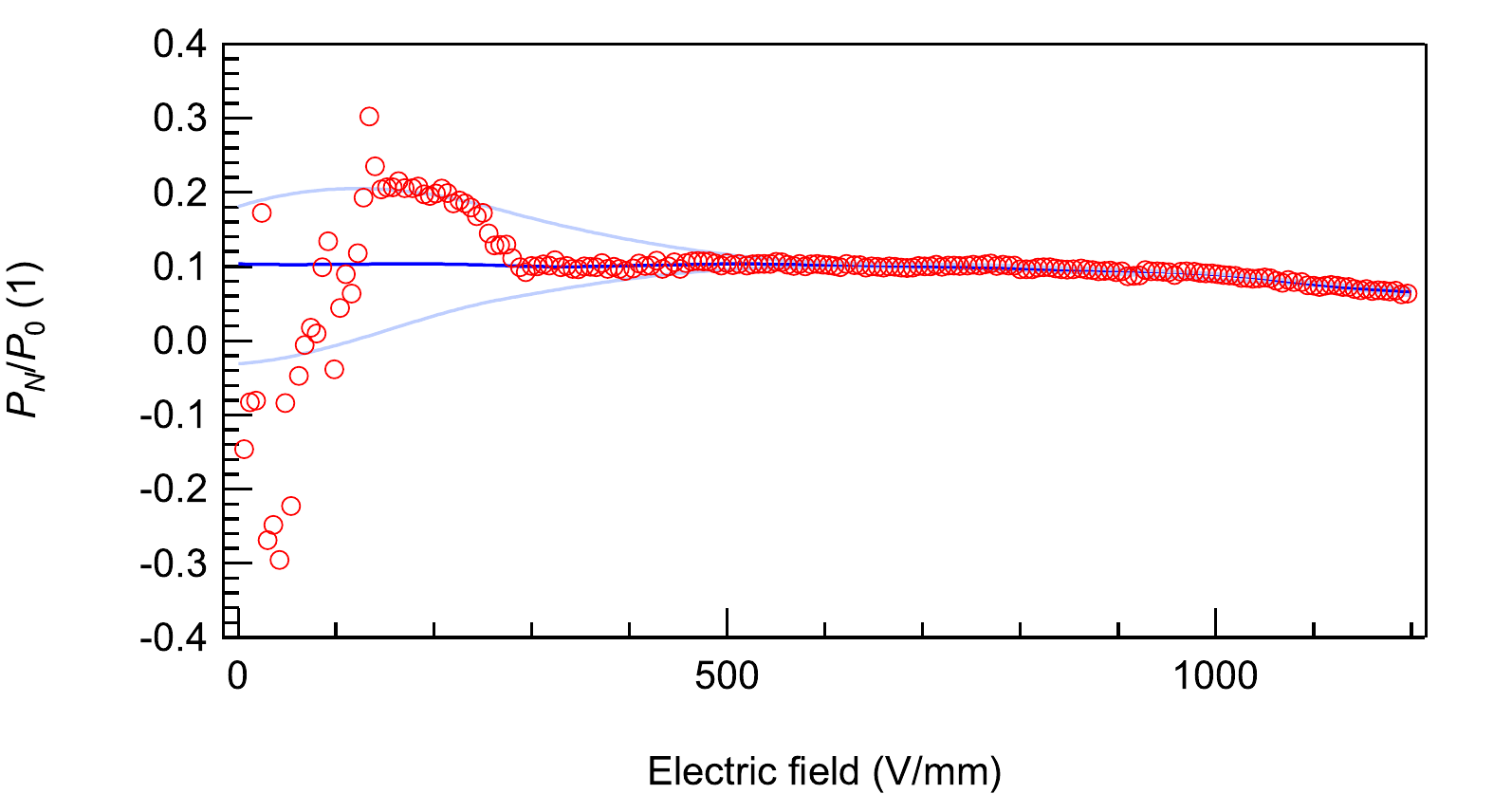}
\caption{Net spontaneous polarization (normalized by the spontaneous polarization $P_0$) as a function of the external field $E_3$. The value $P_N/P_0$ is computed using Eq.~(\ref{eq:PNdef}) where the values of $v_{\rm up}$ and $v_{\rm dn}$ are given by summing the white and black pixels in the domain pattern image [see Fig.~(\ref{fig:results:Domains})].}
\label{fig:results:netpol}
\end{figure}
Further processing of the domain-state image allows the computation of the net spontaneous polarization, which is defined by the formula:
\begin{equation}
	\label{eq:PNdef}
	P_N = P_0\,\left(v_{\rm up} - v_{\rm dn}\right),
\end{equation}
where $v_{\rm up}$ and $v_{\rm dn}$ are the volume fractions of domains with the vector of spontaneous polarization oriented along and against the vector of applied electric field, respectively.
The values of $v_{\rm up}$ and $v_{\rm dn}$ can be easily determined by summing up the white and black pixels in Fig.~(\ref{fig:results:Domains}).
The field dependence of the net spontaneous polarization (normalized by the spontaneous polarization $P_0$) is shown in Fig.~\ref{fig:results:netpol}. 
The red circular markers indicate the measured values.
The thick blue line indicates the median values determined by the non-parametric quantile regression. 
The thin blue lines indicate the first and 10-th deciles, respectively.
It is seen that the accuracy of the DHI method reduces with a decrease in the applied electric field. 
On the other hand, an increase in the magnitude of the external field may influence net spontaneous polarization of the domain pattern.
Therefore, the choice of the magnitude of the applied field represents a delicate procedure of the DHI experiment. 


\begin{figure}
	\centering
	\includegraphics[width=\figAsize]{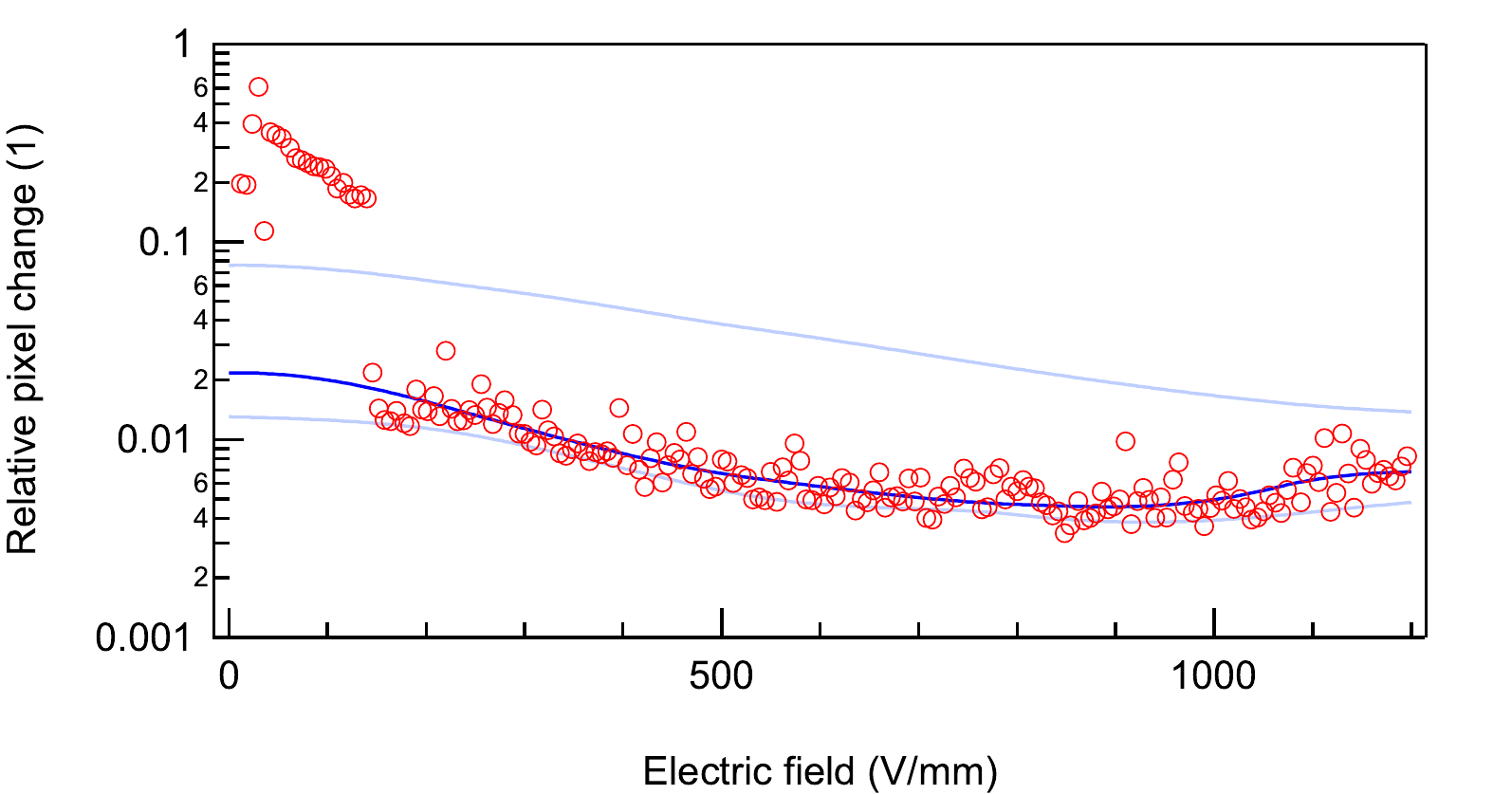}
\caption{Relative change in pixels of the domain pattern image (see Fig.~\ref{fig:results:Domains}) produced by the increase in the magnitude of the external electric field.}
\label{fig:results:relPixchange}
\end{figure}
\begin{figure}
	\centering
	\includegraphics[width=\figAsize]{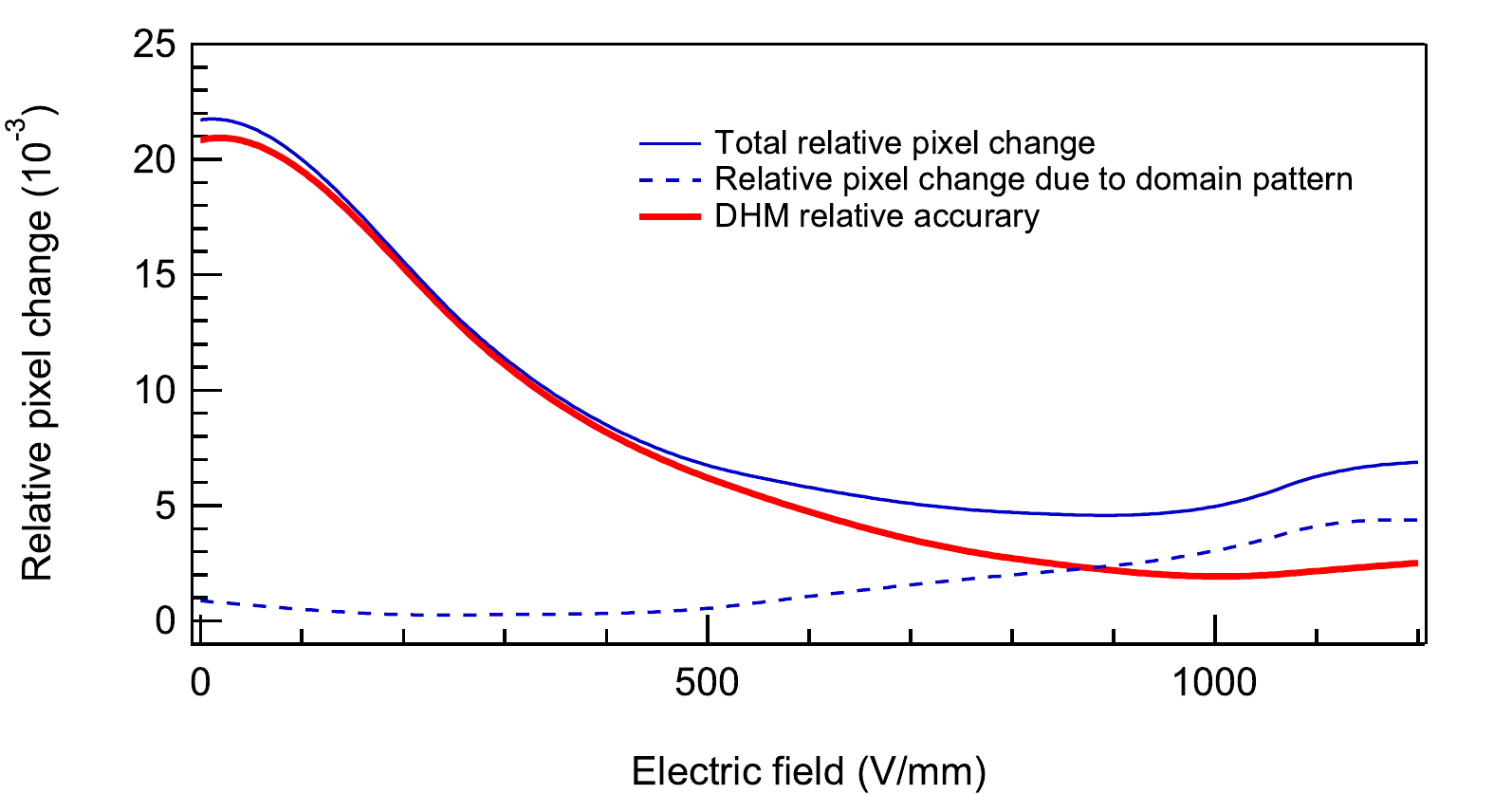}
\caption{Total relative pixel change (thick blue line) as a sum of the relative pixel change due to the domain pattern evolution (dashed blue line) and the DHM relative accuracy (Thick red line). The curves correspond to the measured median values presented in Figs.~\ref{fig:results:netpol} and \ref{fig:results:relPixchange}.}
\label{fig:results:dhmaccuracy}
\end{figure}
In order to determine the optimal amplitude of the external electric field, we have proceeded the following procedure. 
We have performed a series of DHI experiments at different magnitudes of the external field from 0 to 1.2~kV/mm. 
Then, we have analyzed the evolution of the observed domain patterns according to procedure described above. 
Figure~\ref{fig:results:relPixchange} shows the relative change in pixels produced by the increase in the magnitude of the external field.
The total relative change in pixels with an increase in the external field magnitude is controlled by two contributions: (i) by the increased accuracy of the DHI experiment due to greater measured phase difference $\Delta\phi$, and (ii) by the change in the net spontaneous polarization due to the increase in the external electric field.
Since the total relative change in pixels can be measured and the net spontaneous polarization can be reasonably estimated, the relative accuracy of the DHI observation can be estimated.
Results of these estimations are presented in Fig.~\ref{fig:results:dhmaccuracy}.


It should be noted that the computed numerical value of the net spontaneous polarization may be affected by the oversimplified consideration of the piezoelectric effect in Eq.~(\ref{eq:DeltaPhi}).
Indeed, the formula for the phase shift, which is presented in this work by Eq.~(\ref{eq:DeltaPhi}), has been calculated under the consideration of the single-domain ferroelectric sample. 
This is clearly not the case of the particular situation of our experiment. 
Actually, the elastic interaction of the adjacent domain under the application of the external electric field results in ``smearing'' the sharp interface between domains, which is of a sub-nanometer scale in BT single crystal. 
Nevertheless, the numerical phase-field simulations of the similar system~\cite{mokry_identification_2016} indicates that the stray elastic fields in the domain wall regions does not affect the DHM result.


Another advantage of our measurement arrangement is the fact that the wavefront distortion $\Delta\phi$, which is measured using DHM, averages the spatial distribution of the spontaneous polarization over the whole volume of the sample. 
It may happen that the $180^\circ$ domain wall is not perpendicular to the surface of the sample due to arbitrary orientation of the sample or due to the pinning of the domain walls to crystal lattice defects. 
In such a situation, the wavefront distortion is given by the average contribution of the material as the light passes through the thickness of the sample $h$.
As a result, the net spontaneous polarization computed using Eq.~(\ref{eq:PNdef}) corresponds to the volume-averaged value. 
It should be noted that the same interior information on the ferroelectric domain pattern can be provided by means of amplitude-based transmission optical microscopy.
This is a completely different situation as compared e.g. with the domain pattern observation using Piezoresponse Atomic Force Microscope (PAFM), where the domain pattern is monitored from the surface response of the sample.


Additional advantage of the presented method for the observation of domain patterns is that it is not necessary to know the exact values of spontaneous polarization or linear electro-optical coefficients.
It is sufficient to consider that the applied electric field produces the optical contrast in the adjacent domains, which is measurable by the particular implementation of the DHM.
Therefore, our method is suitable for the observation of domain patterns even for those perovskite single crystals, where the exact numerical values of linear electro-optical parameters are not available. 
The numerical value of $r_{333}^T$ can be roughly estimated from Eq.~(\ref{eq:rijk}):
\begin{equation}
	r_{333}^T = g_{11}^T\,\chi_{c}\,P_{0},
\end{equation}
where $g_{11}^T=0.17$~m$^4$C$^{-2}$ is almost temperature independent and universal material parameter for all perovskite materials and the lattice permittivity $\chi_{c}$ and spontaneous polarization $P_{0}$ can be determined using basic dielectric measurements. 
On the other hand, when the complete set of tensor components of $r^T_{ijk}$ is known for the given sample, the precise measurements of the wavefront distortions due to the changes in the refractive index make it possible to distinguish the orientation of the polarization in particular domains. 
Finally, since the refractive index in the perovskite ferroelectric is sensitive to the depolarizing field, the DHM measurements offer a tool for distinguishing the charged and neutral domain walls.


The greatest advantage of this experimental approach stems from the possibility to observe ferroelectric domain patterns at a rather high speed. 
The speed limit of the method is given by the frame rate of the available digital camera and the bandwidth of the data transfer from the camera to a computer processing numerical data. 


There exist two limitations of the DHM based on the electro-optic effect.
The first limitation is the necessity of the external applied field, which may influence the domain pattern or even induce the ferroelectric switching in the sample.
Due to the practical image processing reasons, the minimal required phase shift $\Delta\phi$ to observe the domain pattern is about 0.5~rad. 
Such a phase shift can be achieved in a BT single crystal by applying the external field of the magnitude of about 0.2~kV/mm (see Figs.~\ref{fig:results:netpol} and \ref{fig:results:relPixchange}).
This value is much smaller than the coercive field 1~kV/mm in BT, so that the DHM can be used to study the intermediate switching stages in BT.
The second limitation is the spatial resolution of the DHM method, which is given by the diffraction limit and the wavelength of the used light. 
In the case of the present experiments, it is approximately 500~nm.

\section{Conclusion}

In this Article, we have briefly reviewed the physical principles that allows the observation of ferroelectric domain patterns by means of digital holographic microscopy (DHM).
The principle of the DHM is based on the linear electro-optic effect, which allows to distinguish the anti-parallel domain using transmission interferometric measurements, since the value of the electro-optic coefficient is proportional to the value of the spontaneous polarization, which alters the sign from domain to domain. 
When the external electric field is applied to the ferroelectric polydomain sample, the wavefront of the originally planar optical wave is distorted according to the spatial distribution of spontaneous polarization in the domain pattern. 
The wavefront distortion can be measured using a method called digital holographic interferometry (DHI).


The construction details of our digital holographic microscope (DHM) have been presented. 
Our DHM is based on the Mach-Zehnder interferometer. 
In order to compute the wavefront distortion of the optical wave transmitted through the ferroelectric sample with a domain pattern, two micro-interference patterns (called digital holograms) were captured by our DHM and processed using the angular spectrum method.
The first digital hologram is captured at zero applied electric field and the second one is captured when the external electric field is applied to the sample. 
Using this approach, the wavefront distortion of the optical wave has a useful property, since the anti-parallel domains can be easily distinguished by the sign of the wavefront distortion.
The domain wall regions corresponds to areas on the sample surface, where the wavefront distortion equals zero.


We have demonstrated that using this approach, it is possible to observe the domain patterns in an arbitrary perovskite single crystal sample disregarding the knowledge of numerical values of particular linear electro-optic coefficients.
In addition, the method allows the measurement of the volume fraction of anti-parallel domains in the whole volume of the sample, which is of interest in many applications. 
Finally, the method allows a fast observation of the domain pattern evolution during the physical phenomena such as ferroelectric phase transitions, ferroelectric switching, interactions of domain walls with crystal lattice defects, etc.

\section*{Acknowledgment}

The authors would like to express their sincere gratitude to Jan Ple\v{s}til for his help in the design and construction of the optical cell, and to Tom\'a\v{s} Sluka for many discussion on optical observations of domain walls in barium titanate and for reading the manuscript.
This work was supported by the Czech Science Foundation under Grant GACR 14-32228S.


\end{document}